\documentclass[pra,aps,reprint,nofootinbib,superscriptaddress,floatfix]{revtex4-2}
\usepackage[english]{babel}
\usepackage{jabbrv}
\usepackage[colorlinks=true,linkcolor=blue,citecolor=blue,urlcolor=blue,breaklinks]{hyperref}
\usepackage{amsmath,amssymb,amsfonts}
\usepackage[normalem]{ulem}
\usepackage{graphicx}% Include figure files
\usepackage{dcolumn}% Align table columns on decimal point
\usepackage{bm}% bold math
\usepackage{comment}
\usepackage{physics}
\usepackage{cleveref}
\usepackage{xfrac}
\usepackage{xcolor}
\usepackage{booktabs}
\usepackage{tabularx}
\usepackage{threeparttable}
\usepackage{graphicx}
\usepackage{tikz}
\usepackage{pgfplots}
\usepackage{pgfplotstable}
\pgfplotsset{compat=1.18}

\definecolor{carnelian}{rgb}{0.7, 0.11, 0.11}

\usepackage{float}

\definecolor{darkgreen}{rgb}{0.0, 0.4, 0.0}

\newcommand{\Vpi}{V_{\pi}}

\newcommand{\bps}{\,\mathrm{bits/pulse}}

\begin{document}

\preprint{APS/123-QED}

\title{Poled-fibre phase modulator for efficient high-dimensional quantum measurements} 

\author{N.~Guerrero}
\email{These authors contributed equally to this work}
\affiliation{Departamento de Física, Universidad de Concepción, Concepción, 160-C, Bío Bío, Chile}

\author{N.~Villalba}
\email{These authors contributed equally to this work}
\affiliation{Departamento de Física, Universidad de Concepción, Concepción, 160-C, Bío Bío, Chile}

\author{G.~H.~dos Santos}
\email{fisica.gu@gmail.com}
\affiliation{Departamento de Física, Universidad de Concepción, Concepción, 160-C, Bío Bío, Chile}

\author{C.~J.~Salazar}
\affiliation{Departamento de Física, Universidad de Concepción, Concepción, 160-C, Bío Bío, Chile}

\author{C.~Melo}
\affiliation{
Department of Electrical Engineering, Universidad de Concepción, Edmundo Larenas 219, Concepción 4030000, Chile}

\author{F.~Castillo}
\affiliation{
Department of Electrical Engineering, Universidad de Concepción, Edmundo Larenas 219, Concepción 4030000, Chile}

\author{J.~Cari\~ne}
\affiliation{Departamento de Ingenier\'{\i}a El\'ectrica, Universidad Cat\'olica de la Sant\'{\i}sima Concepci\'on, Concepci\'on, Chile}

\author{G.~B.~Xavier}
\affiliation{Institutionen för Systemteknik, Linköpings Universitet, 581 83, Linköping, Sweden}

\author{E.~S.~G\'omez}
\affiliation{Departamento de Física, Universidad de Concepción, Concepción, 160-C, Bío Bío, Chile}
\affiliation{Millennium Institute for Research in Optics, Universidad de Concepción, Concepción, 160-C, Bío Bío, Chile}

\author{S.~P.~Walborn}
\affiliation{Departamento de Física, Universidad de Concepción, Concepción, 160-C, Bío Bío, Chile}
\affiliation{Millennium Institute for Research in Optics, Universidad de Concepción, Concepción, 160-C, Bío Bío, Chile}

\author{J.~Pereira}
\affiliation{Fibre Optics, RISE-Research Institutes of Sweden, Electrum 236, 16440, Kista, Sweden}

\author{G.~Saavedra}
\affiliation{
Department of Electrical Engineering, Universidad de Concepción, Edmundo Larenas 219, Concepción 4030000, Chile}

\author{G.~Lima}
\email{glima@udec.cl}
\affiliation{Departamento de Física, Universidad de Concepción, Concepción, 160-C, Bío Bío, Chile}
\affiliation{Millennium Institute for Research in Optics, Universidad de Concepción, Concepción, 160-C, Bío Bío, Chile}

\date{\today}

\begin{abstract}
Efficient detection of quantum states underpins advanced device-independent quantum-information protocols that provide the ultimate level of security for tasks including quantum random number generation and quantum key distribution (QKD). High-dimensional encoding is a natural route to boost the performance of such protocols, offering enhanced noise resilience and higher information capacity, yet their practical implementation remains challenging. A key experimental bottleneck in higher dimensions is the typical need of active modulators for basis selection, which incur substantial optical losses and polarization-sensitive operation. Poled optical fiber phase modulators (PFPMs) are a fiber-native electro-optic technology that naturally addresses these challenges, combining sub-dB insertion loss, intrinsic polarization independence, and direct compatibility with standard telecommunications fiber. Here we report the first use of a PFPM for active quantum-state measurements in a fully fiber-integrated platform. Basis selection in our receiver for four-dimensional qudits is achieved using a single PFPM, substantially simplifying the receiver architecture. As a benchmark, we perform a four-dimensional QKD session and obtain a finite secret-key rate per pulse that, to the best of our knowledge, surpasses all previously reported QKD demonstrations. Our results establish poled-fiber electro-optic modulation as a broadly applicable platform for high-efficiency detection in fiber-integrated quantum information processing.
\end{abstract}

\maketitle

\section*{Introduction}

\begin{figure*}[t]
\centering
\includegraphics[width=0.75\linewidth]{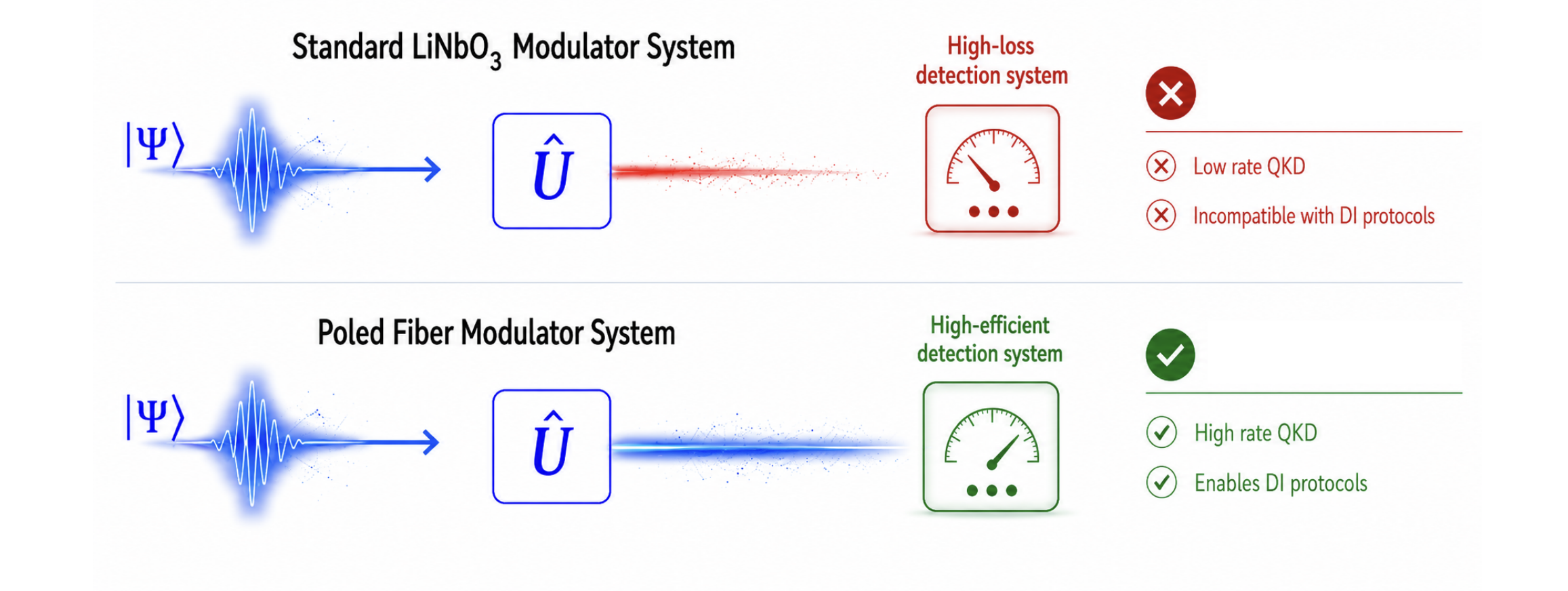}
\caption{\textbf{Effect of receiver loss in quantum photonic experiments.} Top: a conventional receiver based on a commercial, fiber-pigtailed LiNbO$_3$ phase modulator ($3\,$dB insertion loss) implements the basis-selection unitary $U$, pushing the overall measurement efficiency below the threshold required for DI quantum information experiments or severe limiting the rate attainable in QKD. Bottom: a receiver based on a poled-fiber phase modulator implements the same unitary at sub-dB insertion loss, constituting the architecture demonstrated in this work.}
\label{fig:certification}
\end{figure*}

Photonic quantum information has progressed from proof-of-principle demonstrations of quantum cryptography to increasingly complex protocols in communication, sensing, and computation~\cite{bennett1984bb84,Pirandola2020Advances,wehner2018quantum}. Encoding information in photonic states of dimension $d>2$, the so-called qudits, can increase the system information capacity, improve tolerance to noise, and provide access to stronger forms of non-classical correlations~\cite{BechmannTittel2000LargerAlphabets, Cerf2002DLevelQKD,Cozzolino2019HDCommunication, Erhard2020HDEntanglement,malik2026highdimensionalquantumphotonicsroadmap,Mirhosseini2015TwistedLight,Islam2017TimeBinQudits,Canas2017,Zahidy2024DeployedHDQKD,Xavier2025}. In practice, however, the benefits of high-dimensional encoding are achieved only when the receiver performs the measurement at high fidelity and low loss, like it is the case with qubit-based communicating systems. Receiver loss directly suppresses the secret-key rate in quantum key distribution (QKD)~\cite{Sheridan2010QuditSecurity} and opens loopholes in device-independent (DI) protocols~\cite{Acin2007DIQKD,Vertesi2010DetectionLoopholeQudits,Pironio2010Randomness,Argillander2025MDIQRNG} (see Fig.~\ref{fig:certification}). Detection efficiencies sufficient high for high-dimensional DI quantum information processing have been achieved for static, single-outcome measurements \cite{Hu2022HDBellTest}. However, a wide class of protocols requires active measurement-basis selection and multiple-outcome measurements, in which a single detection event determines which of the $d$ outcomes of the selected basis occurred. Active switching inherently introduces several implementation challenges. The most significant is the additional insertion loss associated with the switching elements, followed by limitations arising from finite switching speeds, modulation depth, and, finally, the need to manage other photonic degrees of freedom, such as polarisation. For example, commercial fiber-pigtailed LiNbO$_3$ modulators can operate at GHz, but they introduce an extra $3\,$dB of loss with a strong polarisation dependence \cite{Carine2020,Taddei2021}.

Previous routes to lower-loss active measurements have different trade offs: spatial light modulators and digital micro-mirror devices are flexible but limited to switching rates up to a few kHz~\cite{Etcheverry2013,Mirhosseini2015TwistedLight}; multi-plane light converters are efficient for static measurements but difficult to reconfigure dynamically~\cite{Lib2025MPLC}; piezoelectric fiber stretchers have sub-dB insertion loss but switching is limited to tens of kHz~\cite{Nakamura:23}; and integrated photonic circuits, although recently reaching gigahertz-rate operation, still suffer from relevant fiber-coupling losses~\cite{bernardi_gigahertz-rate_2026}. 

Electro-optic modulation in poled silica fiber is a promising technology that can provide simultaneous low loss and fast response time, thus solving the efficiency issue for high-dimensional device-independent applications. Glass has a low non-linear coefficient for electro-optical modulation, but through thermal or optical poling, a higher effective $\chi^{(3)}$ is induced directly in the fiber itself~\cite{Myers1991PoledSilica,Kazansky1997FiberPoling}. Poled-fiber devices have been successfully operated in classical applications such as fiber-integrated interferometry and switching for over two decades~\cite{Fokine2002FiberMZI,SpegelLexne2026Buffer}. Because the optical mode never leaves the fiber, the fiber-to-crystal coupling responsible for most of the loss in commercial pigtailed devices is eliminated, and the modulation is furthermore intrinsically polarization independent. Nevertheless, the potential of this mature and low-loss technology for quantum-information tasks has remained largely unexplored. 

Here we report the first application of a poled optical fiber phase modulator (PFPM) in a quantum-information experiment. The PFPM has a sub-dB insertion loss of 0.45~dB, intrinsic polarization independence \cite{Tarasenko:06}, and direct compatibility with standard telecommunications fibre, enabling an efficient fibre-integrated architecture for the detection of high-dimensional photonic states. Specifically, used as the active basis-selection element of a four-dimensional quantum receiver, the PFPM allowed the insertion loss of the detection system to be reduced from $5.76,$dB, obtained using LiNbO$_3$ phase modulators, to $2.79,$dB. The resulting detection efficiency of the system then reaches an overall value of $\approx 44\%$, including the detection efficiencies of four superconducting nanowire single-photon detectors (SNSPDs). To demonstrate the potential of this technology for quantum information processing, we use our new detection scheme for implementing a four-dimensional decoy-state QKD session that, up to our knowledge, yields the highest finite secret-key rate per pulse reported to date for any QKD demonstration, at the typical reference channel attenuation level of 10 dB. The broad relevance of poled-fiber electro-optic modulation is further illustrated by the companion submission~\cite{SpegelLexne2026Buffer}, in which a PFPM serves as the active switching element of a quantum packet buffer. Together, the two results establish poled-fiber electro-optic modulation as a versatile, ultra-low-loss, and active technology for fiber-integrated quantum information, with applications spanning high-dimensional QKD, device-independent protocols, and quantum memories.

\section*{Results}

\subsection*{PFPM fabrication and characterization}
\label{sec:pfpm}

\begin{figure}[t]
\includegraphics[width=0.95\linewidth]{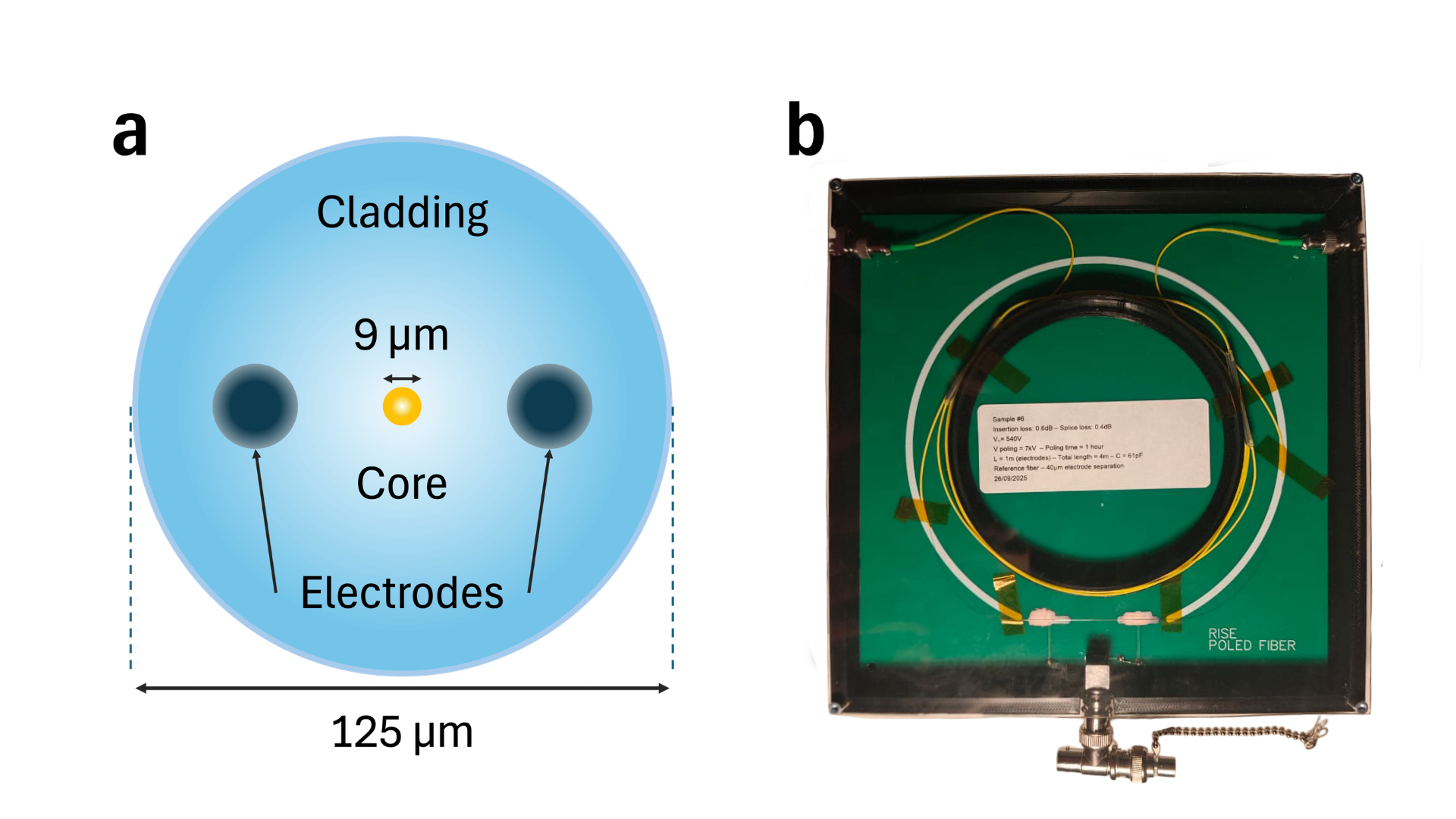}
\caption{a) Cross-section diagram of the poled-fiber phase modulator. Two longitudinal electrodes are inserted in the cladding equidistant from the single-mode core. b) Finished and packaged PFPM.}
\label{fig:pfpm-device}
\end{figure}

A poled-fiber phase modulator is an all-fiber electro-optic device in which metallic electrodes embedded longitudinally through the cladding of a single-mode optical fiber enable voltage-controlled phase modulation (Fig.~\ref{fig:pfpm-device}(a)). Because the optical mode remains guided in the fiber core, the device can be directly spliced into standard fiber systems, avoiding the free-space-to-fiber or crystal-to-fiber coupling losses associated with standard electro-optic modulators~\cite{Myers1991PoledSilica,Kazansky1997FiberPoling,Fokine2002FiberMZI}. 

\begin{figure*}[t]
    \centering
    \includegraphics[width=0.95\linewidth]{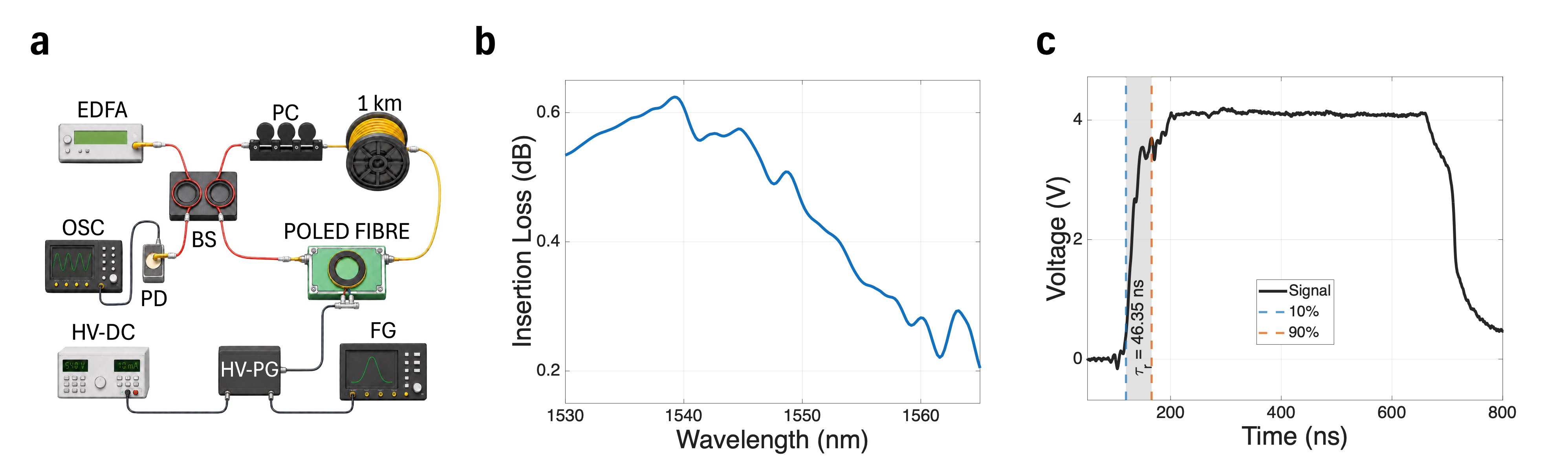}
    \caption{Characterization of the poled-fiber phase modulator.
    \textbf{(a)}, Experimental configuration used for the optical and temporal characterization. Spontaneous amplified emission from an EDFA is injected into the fiber system containing the poled-fiber modulator.
    \textbf{(b)}, Wavelength-dependent insertion loss obtained from measurements with an optical spectrum analyzer.
    \textbf{(c)}, Photodiode waveform measured at the \(\pi\)-phase operating point, including the 10--90\% rise-time estimation. BS: Fiber beamsplitter; EDFA: Erbium Doped Fiber Amplifier; FG: Function generator; HV-DC: High-voltage direct current power supply; HV-PG: High-voltage pulse generator; OSC: Oscilloscope; PC: Manual polarization controller; PD: Photodiode.}
    \label{fig:pfpm-characterization}
\end{figure*}

The PFPM was fabricated by drilling two longitudinal holes in a standard single-mode fiber preform, drawing the preform. After cutting the fiber in smaller sections ($\approx3\,$m), and filling the holes with molten BiSn alloy with pressure, while keeping both ends free of metal. After solidification, the resulting $1\,$m-long metallic electrodes were exposed by polishing the fiber from the side and connected to $20\,\mu$m tungsten wires (Fig.~\ref{fig:pfpm-device}(b)). Since no electrical current can flow between the electrodes, the device behaves as a capacitor with a measured capacitance of $61\,$pF. To overcome the low optical nonlinearity of glass, the fibre undergoes a poling step in which 532 nm light from a Nd:YAG laser is launched into the core under a simultaneously applied high voltage of $\approx$ 7 kV. This rearranges the internal charge distribution and imprints a permanent, long-lived electric field in the glass \cite{Camara15}.

The PFPM was characterized through a wavelength-resolved transmission measurement and a temporal characterization resorting to a fiber Sagnac interferometer (Fig.~\ref{fig:pfpm-characterization}(a)). The optical characterization in Fig.~\ref{fig:pfpm-characterization}(b) uses an erbium-doped fiber amplifier (EDFA) as a broadband amplified-spontaneous-emission (ASE) source between 1528 and 1568 nm. The ASE light is split by a 50/50 fiber coupler into a fiber-based Sagnac loop containing a polarization controller, a $1\,$km fiber spool to temporally separate the clockwise and counterclockwise pulses, and the PFPM, and is finally detected on a photodiode. The PFPM is driven by a high-voltage DC source (HV-DC) gated by a high-voltage pulse generator (HV-PG), itself triggered by a TTL signal of $400\,$ns at $1\,$kHz from a function generator. The wavelength-resolved insertion loss is measured on an optical spectrum analyzer with the Sagnac loop open. The repetition rate is presently limited by the high-voltage drive electronics rather than by the PFPM itself. Because the PFPM is predominantly capacitive, the current limitation is set by the high-voltage driver and its $50\,\Omega$ termination rather than by dissipation in the fibre device itself. The $10\,\mathrm{mA}$ current limit of the HV-DC supply therefore imposes a practical trade-off between TTL gate width and repetition rate, since wide gate pulses can drive the electronics into over-current protection. For $40\,\mathrm{ns}$ electrical pulses, stable operation was experimentally verified up to $100\,\mathrm{kHz}$ drawing a current of $8.5\, \mathrm{mA}$. In principle, however, the same driving electronics can support repetition rates up to a few $\mathrm{MHz}$ provided that the current load remains below the protection threshold.

The insertion loss in Fig.~\ref{fig:pfpm-characterization}(b) was extracted by subtracting, in dB, the spectrum measured after the PFPM from a reference spectrum recorded before the device with the same optical-spectrum-analyzer settings. This gives an average loss of $0.45\,$dB over $1530$--$1565\,$nm, with only weak spectral variation across the telecom C-band. The temporal response at the half-wave operating point [Fig.~\ref{fig:pfpm-characterization}(c)] yields $V_\pi = 540\,$V and a 10--90\% rise time of $\tau_r = 46.35\,$ns.

%---------------------------------------------------------------
\subsection*{The PFPM for quantum information processing}
\label{sec:setup}
%---------------------------------------------------------------

\begin{figure*}[t]
\centering
\includegraphics[width=0.8\linewidth]{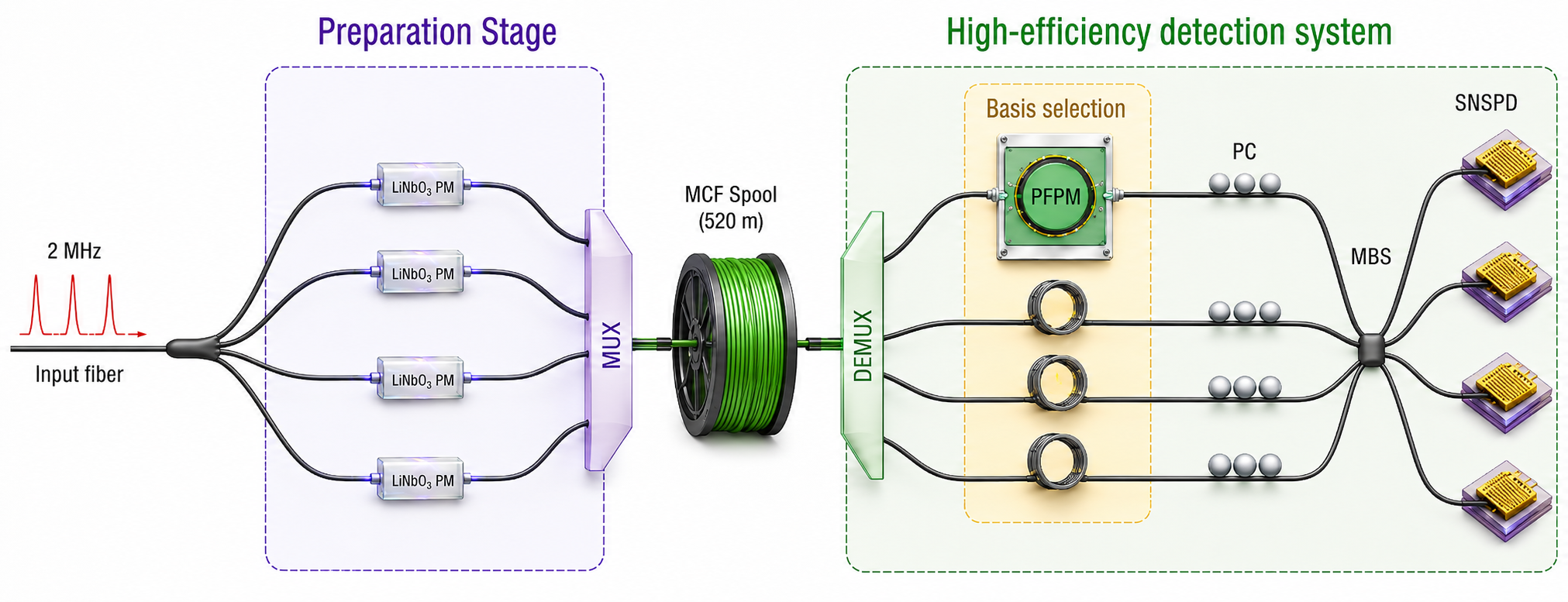}
\caption{Schematics of the four-dimensional HD-QKD system used. A weak coherent pulse is distributed among four single-mode fibers through a four core multicore beam splitter (4C-MBS)~\cite{dosSantos2026} and sent to the Preparation Stage, where four LiNbO$_3$ phase-modulators generate the QKD qudit states. The photons are then multiplexed into a $520\,$m spool of a 4C multicore fiber, which constitutes the transmission channel. At Bob, the photons are routed to the high-efficiency detection system, within which a single PFPM (yellow Basis Selection box) realises the active basis switching. Four polarization controllers ensure path indistinguishability before a second 4C-MBS performs the high-dimensional projection, with the detection events recorded by four SNSPDs.}
\label{fig:setup}
\end{figure*}

To evaluate the performance of the PFPM for quantum information processing, we integrate it into a four-dimensional ($d=4$) phase-coding HD-QKD system~\cite{dosSantos2026}, in which the qudit logical states are encoded in terms of the path/core modes available for photon propagation through a four-core multicore fiber (4C-MCF): a single photon propagating over path/core $k\in\{0,1,2,3\}$ is identified with the basis state $\ket{k}$. The pulse generation, decoy-state implementation, synchronization electronics and state-preparation hardware follow the architecture described in Ref.~\cite{dosSantos2026}. A schematic overview of the experiment is shown in Fig.~\ref{fig:setup}.

At Alice, a weak coherent pulse at $1550\,\mathrm{nm}$ is distributed among four single-mode fibers by a four-core multicore beam splitter (4C-MBS), creating an equal-amplitude path superposition. The pulse then enter the Preparation Stage, where a set of LiNbO$_3$ phase modulators imprint a phase pattern $\vec{\phi}^A$ between the four paths, producing initial states of the form given by $\ket{\Psi(\vec{\phi}^A)} = \tfrac{1}{2}\sum_{k=0}^{3} e^{i\phi_k^A}\,\ket{k}$. The complete list of the initial states used is given in Methods. Note that Alice does not require low-loss modulators, since insertion loss at the transmitter must be further increased to reach the extreme weak coherent states used in quantum information processing. The resulting coherent state is subsequently multiplexed into a 4C-MCF through a single-mode fiber (SMF) to MCF multiplexer (MUX).

Alice and Bob are connected through a $520\,$m homogeneous 4C-MCF spool spliced to the transmitter and receiver stations. The spool contributes to $\lesssim 2.0\,$dB of loss and the relative phase between cores remains essentially constant over the $100\,$ms probe-lock-guard integration time window~\cite{dosSantos2026}. The total channel attenuation is set to $10\,$dB through variable attenuators, providing a common reference point for comparison with recent finite-key QKD demonstrations~\cite{Islam2017TimeBinQudits,Zahidy2024DeployedHDQKD,dosSantos2026,Li2023HighRate110Mbps,Zhang2025SPS,DaLio2021PathMCF}.

At Bob's receiver, a four-dimensional measurement in one of two bases is performed. The demultiplexer (DEMUX) maps each core into an individual single-mode fiber arm connected to the Basis Selection Stage. The first arm hosts the active PFPM, while the remaining three arms carry only short, length-matched fiber patch cords whose static phases are absorbed into Bob's per-block phase calibration. A manual polarization controller is placed on each of the four arms, following the PFPM and the three patch cords, to ensure polarization-mode indistinguishability of the four arms. After the basis-selection stage, multi-path interference at a second 4C-MBS implements a projective measurement in one of the two QKD bases actively chosen by the PFPM, and each of the four output ports is fiber-coupled to a superconducting nanowire single-photon detector (SNSPD) with system detection efficiency $\eta_d = 85\%$ and dark-count probability of $\approx 1{,}3 \times 10^{-5}$.

The two MUBs realised at Bob have their eingenvectors defined in terms of equal-amplitude superpositions of the four cores states $\ket{k}$. Since our receiver is based on a multicore beamsplitter, it can be modelled using the Sylvester--Hadamard formalism for any arbitrary dimension $d$~\cite{Horadam2007,Carine2020}. In this case, switching between two MUBs typically requires phase patterns acting on multiple arms, which motivates a bank of $d$ independent phase modulators (one per arm) and was the architecture of our previous LiNbO$_3$-based receiver~\cite{dosSantos2026}. A structural property of the four-dimensional case, however, makes a single phase modulator sufficient: the two MUB phase patterns at Bob differ only by a $\pi$ phase on one arm~\cite{Canas2017,Carine2020,dosSantos2026}. The PFPM applies this phase at $V = \Vpi = 540\,$V, realising the $\mathcal{X}$ projection, while $V = 0$ realises the $\mathcal{Z}$ projection. Thus, the remaining modulators would be redundant for basis switching in QKD. This statement is formally demonstrated in Methods. The key fact that the PFPM has a sub-dB insertion loss does not compromise the quality of the final interference observed, and consequently barely affect the resulting QBER, which would be not the case if only one LiNbO$_3$ had been used.

%===============================================================
\subsection*{PFPM-based HD-QKD}
\label{sec:performance}
% ===============================================================

Stable interferometric operation of the receiver is enforced by the probe--lock--guard control loop introduced in Ref.~\cite{dosSantos2026} and implemented with Alice's PMs.  This relax the stress under the PFPM at the receiver, which now work for QKD acquisition only when the phase-stability metric satisfies $\mathrm{QBER}<2.5\%$. Under this condition the accepted acquisition windows yield mean QBERs of $E_{\mu \mathcal{Z}}=2.83\%$ and $E_{\mu \mathcal{X}}=4.52\%$, both well below the $18.9\%$ coherent-attack threshold for $d=4$~\cite{Sheridan2010QuditSecurity}. The residual asymmetry between the two settings is consistent with the calibration uncertainty of the PFPM half-wave point: a small deviation from the exact $\pi$ phase affects only the driven $\mathcal{X}$ projection, whereas the $\mathcal{Z}$ projection is obtained with the PFPM unbiased.

Using the measured gains and error rates in $\mathcal{Z}$ and $\mathcal{X}$ basis (see Methods for the acquisition protocol and the complete set of recorded parameters), we evaluate the finite secret-key fraction $R$ expected for a four-dimensional decoy-state implementation. At the reference attenuation of $10\,$dB the calculation yields
\begin{equation}
R_{\mathrm{finite}}(10\,\mathrm{dB}) \;=\; 1.06 \times 10^{-2}\bps.
\label{eq:final}
\end{equation}

\begin{figure}[t]
\centering
\includegraphics[width=0.95\linewidth]{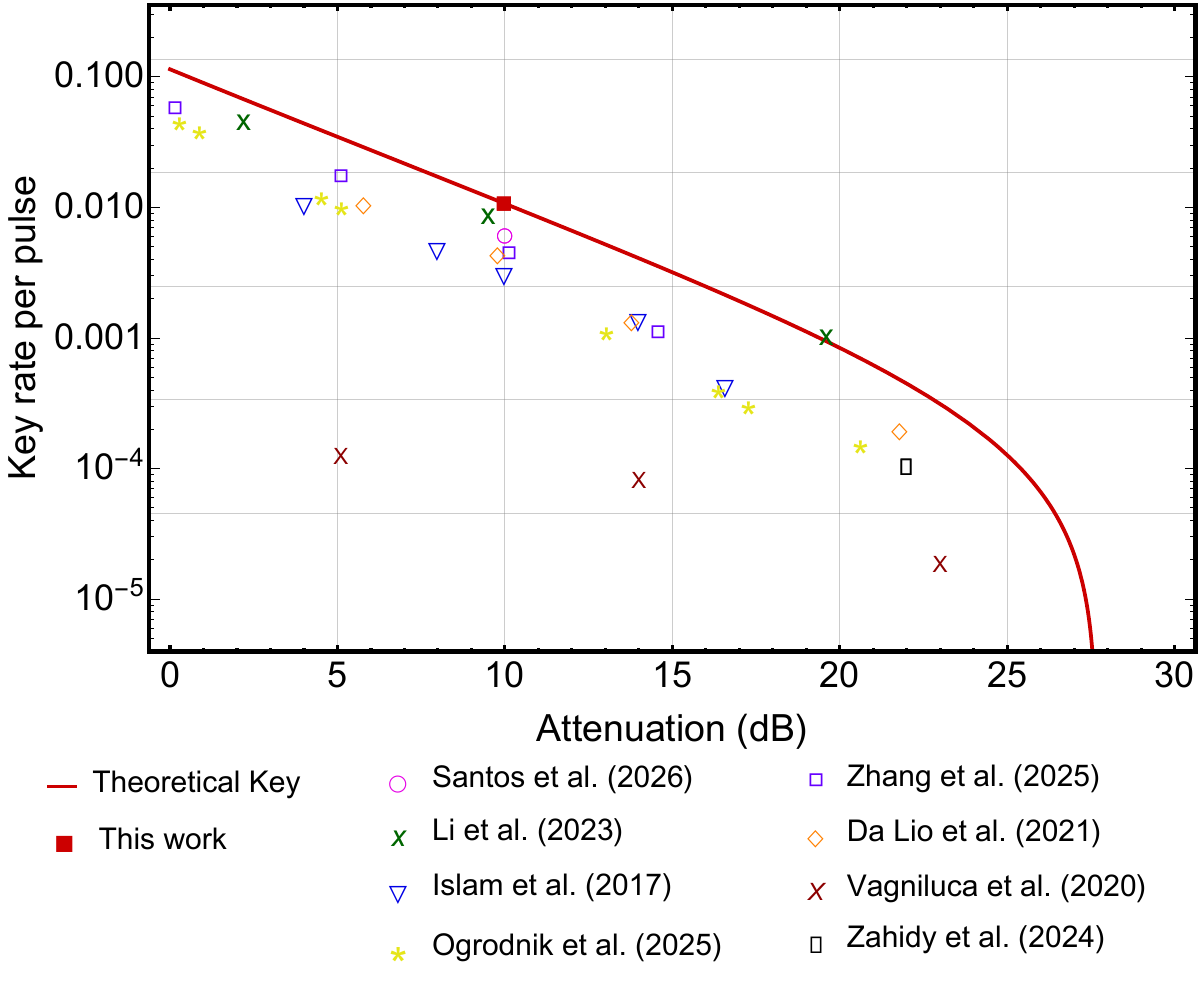}
\caption{Finite secret-key fraction (bits per pulse) as a function of channel attenuation, compared with recent leading QKD demonstrations.}
\label{fig:keyrate}
\end{figure}

Figure~\ref{fig:keyrate} places this result in the context of recent leading QKD demonstrations across different encoding platforms. At the common $10\,$dB reference attenuation, the present rate exceeds all previously reported QKD experiments, including the best $d=2$ implementations~\cite{Li2023HighRate110Mbps,Zhang2025SPS}. Up to our knowledge, this is the first time a high-dimensional QKD implementation achieves competitive performance with what has been achieved already by the top leading bi-dimensional QKD implementations.

\section*{Discussion}
\label{sec:discussion}

Recently we demonstrated a 4D-QKD protocol over a deployed MCF network, where Bob receiver was built with four LiNbO$_3$ phase modulators instead of a single PFPM~\cite{dosSantos2026}. Comparing that to the present work, the total receiver insertion loss improved from $5.76\,$dB to $2.79\,$dB (Table~\ref{tab:bob-budget}), corresponding to an increase of the receiver overall transmission from $26.5\%$ to $52.6\%$. The corresponding finite secret-key rate improves by a factor of $\sim 1.7$ at $10\,$dB, going from $6.18\times 10^{-3}$ to $1.06\times 10^{-2}\,$bits/pulse, with the main improvement coming from the use of the PFPM.

\begin{table}[t]
\caption{Bob insertion-loss budget, by component. The two contributions that changed between the architectures are the active phase modulator (LiNbO$_3$ to PFPM), and the number of required PC stages (eight PCs to four PCs, with the four LiNbO$_3$-internal input polarizers also eliminated since the PFPM is polarization-independent).}
\centering
\begin{tabular}{lcc}
\toprule
Component & LiNbO$_3$~\cite{dosSantos2026} & PFPM (this work) \\
\midrule
Phase modulator      & $3.22$~dB  & $0.45$~dB \\
PC stages + conectors      & $0.30$~dB  & $0.10$~dB  \\
MCF mux + demux      & $1.43$~dB  & $1.43$~dB  \\
4C-MBS      & $0.62$~dB  & $0.62$~dB  \\
Splices   & $0.19$~dB  & $0.19$~dB \\
\midrule
\textbf{Total}       & \textbf{5.76~dB} & \textbf{2.79~dB} \\
\bottomrule
\end{tabular}
\label{tab:bob-budget}
\end{table}

As conceptually illustrated by Fig.~\ref{fig:certification}, the PFPM-based receiver addresses the receiver-loss bottleneck of high-dimensional DI photonic protocols. Doubling the receiver transmission, as shown by Table~\ref{tab:bob-budget}, closes a substantial fraction of the gap between current HD photonic receivers and the efficiency regime required by device-independent protocols; the impact compounds in multi-photon Bell tests and multi-party DI protocols. For instance, the $44\%$ overall detection efficiency of the present receiver already lies above the $43\%$ asymmetric threshold identified for the $I_{3322}$ Bell inequality when one party has perfect detection efficiency~\cite{Brunner2007AsymmetricBell}, a regime relevant to atom--photon hybrid Bell tests where the matter party is detected with near-unit efficiency. 

The residual loss is now dominated by the MCF mux/demux stage ($1.43\,$dB) and the 4C-MBS ($0.62\,$dB), both fiber-integrated components amenable to further loss reduction without sacrificing fiber compatibility or basis-switching speed. The reduced receiver loss broadens the operating regime accessible to the platform: although high-dimensional QKD is at a disadvantage in high-loss regimes, where the relative weight of dark counts grows faster with attenuation than in $d=2$ implementations, the present loss budget supports competitive finite-key rates up to approximately $20\,$dB of channel attenuation ($\approx 100\,$km), as projected by the curve in Fig.~\ref{fig:keyrate}.

Together with the companion paper demonstrating a fiber-integrated quantum packet buffer~\cite{SpegelLexne2026Buffer}, the PFPM emerges as a versatile electro-optic device for fiber-based quantum networks, spanning low-loss state preparation, high-fidelity high-efficient quantum measurements, and active routing and buffering. Beyond specific protocols, low-loss fiber-integrated electro-optic control is directly applicable to measurement-based photonic processing, feed-forward operations and the broader class of fiber-integrated quantum receivers and memories. Our experimental results position poled-fiber electro-optics as a promising building block for scalable, fiber-integrated quantum photonic technologies.

\section*{Methods}
\label{sec:methods}

\subsection*{Finite-key analysis}

We use a prepare-and-measure HD-BB84 protocol with phase coding in $d=4$ spatial modes and the standard vacuum$+$weak two-decoy scheme~\cite{Lo2005,Ma2005}. 

The finite-key analysis follows the vacuum$+$weak formalism of Zhang et al.~\cite{Zhang2017}, extended to $d$ dimensions. For a sample of $N$ sent pulses and security parameters $\epsilon_{\mathrm{sec}}$ and $\epsilon_{\mathrm{cor}}$, the extractable secret-key fraction in $d$-dimensional decoy-state QKD is bounded by
\begin{align}
R \;\ge\; \frac{1}{N}\Bigl[\,
& M_z^{0,L}\,\log_2 d
\;+\; M_z^{1,L}\!\left(\log_2 d - H_d(e_1^U)\right) \nonumber\\
& \;-\; M_{\mu z}\, H_d(E_{\mu z})\,f(E_{\mu z}) \nonumber\\
& \;-\; 6\log_2(21/\epsilon_{\mathrm{sec}})
\;-\; \log_2(2/\epsilon_{\mathrm{cor}}) \,\Bigr],
\label{eq:Rfinite}
\end{align}
where $M_z^{0,L}$ and $M_z^{1,L}$ are Chernoff lower bounds on the vacuum and single-photon detection counts in the $\mathcal{Z}$ basis, estimated from the decoy measurements. The parameter $e_1^U$ is the random-sampling upper bound on the single-photon error rate. $M_{\mu z}$ and $E_{\mu z}$ are the measured gain and QBER for the signal intensity $\mu$ in $\mathcal{Z}$. $f(E_{\mu z})$ is the error-correction inefficiency, and $H_d(x) = -x\log_2[x/(d-1)] - (1-x)\log_2(1-x)$ is the $d$-ary Shannon entropy. The explicit Chernoff/Hoeffding constants and the decoy-bound expressions used to evaluate Eq.~\eqref{eq:Rfinite} are taken from Ref.~\cite{dosSantos2026}. Nevertheless, naturally all the experimental recorded parameters listed in Table~\ref{tab:finite} change between the two implementations.

\subsection*{Data acquisition and finite-key benchmark at $10\,$dB}

The gains and QBERs that enter the finite-key bound were acquired in a block-wise sequence aligned with the probe--lock--guard control loop. Each cycle consists of a $100\,$ms lock window, during which Alice's four LiNbO$_3$ phase modulators apply the reference pattern that drives the closed-loop interferometric stabilization~\cite{dosSantos2026}, and the PFPM is held at the drive voltage corresponding to the basis being characterized ($V=0$ for $\mathcal{Z}$, $V=V_\pi$ for $\mathcal{X}$). This stabilization procedure is followed by a $100\,$ms acquisition window during which the PFPM voltage is chosen to be $V=0$ or $V=V_\pi$, while Alice prepares one of the QKD states. The resulting gains and QBER for that all basis are recorded. Successive $\mathcal{Z}$ and $\mathcal{X}$ blocks are interleaved to accumulate the statistics that enter Eq.~\eqref{eq:Rfinite}. In this acquisition procedure the PFPM is not required to switch at the optical clock rate, as it applies the voltage after path stabilization has been accomplished. Nonetheless, its $46\,$ns electrical response is compatible with per-pulse active basis selection.

Over the full session we accumulated $N = 2.75 \times 10^9$ sent pulses across the two receiver settings, with optimized decoy intensities $(\mu, \nu) = (0.640, 0.092)$, sending probabilities $(p_\mu, p_\nu, p_0) = (0.744, 0.214, 0.042)$ and basis-selection probability $p_z = 0.949$. The security parameters were set to $\epsilon_{\mathrm{sec}} = 2.92 \times 10^{-8}$, $\epsilon_{\mathrm{cor}} = 10^{-15}$ and overall failure probability $\epsilon = 10^{-10}$. Under these conditions, Eq.~\eqref{eq:Rfinite} yields a finite-key length of $K = 2.92 \times 10^{7}\,$bits for the full accepted data set, corresponding to $R_{\mathrm{finite}}(10\,\mathrm{dB}) = 1.06\times 10^{-2}\,\bps$. The detection statistics and the final key length are summarized in Table~\ref{tab:finite}.

\begin{table}[t]
\centering
\caption{Measured detection and error statistics, and the corresponding finite-key bound at $10\,$dB of channel loss. The decoy intensities and probabilities have been jointly optimized for the receiver loss of $2.79\,$dB.}
\label{tab:finite}
\begin{tabular}{lc}
\toprule
Parameter & Value \\
\midrule
Channel loss               & $10\,$dB \\
Receiver loss (Bob)        & $2.79\,$dB \\
SNSPD efficiency $\eta_d$  & $85\%$ \\
\midrule
$N$                        & $2.75\times10^{9}$ \\
$M_z$                      & $5.44\times10^{7}$ \\
$M_{\mu z}$                & $5.21\times10^{7}$ \\
$M_{\nu z}$                & $2.22\times10^{6}$ \\
$M_{\mu x}$                & $1.44\times10^{5}$ \\
$M_{\nu x}$                & $6.49\times10^{4}$ \\
$\overline{E}_{\mu z}$     & $2.83\%$ \\
$\overline{E}_{\mu x}$     & $4.52\%$ \\
$M_0^L$                    & $5.32\times10^{4}$ \\
$M_1^L$                    & $2.66\times10^{7}$ \\
$e_1^U$                    & $6.14\%$ \\
$f(E_\mu)$                 & $1.05$ \\
\midrule
$K$ (accepted finite-key data set)  & $2.92\times10^{7}$ bits \\
$R_{\mathrm{finite}}$      & $\bm{1.06\times10^{-2}\,\bps}$ \\
\bottomrule
\end{tabular}
\end{table}

\subsection*{Single-modulator basis projection}

The eight states prepared by Alice in the protocol are listed in Table~\ref{tab:states}. The receiver implements projective measurements by applying a diagonal phase operation before a fixed four-port Sylvester--Hadamard multiport that is well-described by a normalized Hadamard transformation \cite{Carine2020}:
\[
H_4=\frac{1}{2}
\begin{pmatrix}
1&1&1&1\\
1&-1&1&-1\\
1&1&-1&-1\\
1&-1&-1&1
\end{pmatrix}
\]
 
 In the absence of an additional phase modulation, the analyzer projects onto the Hadamard basis. If a phase shift is applied before the multiport, the relevant overlap matrix between the two analyzer bases is
\[
U(D)=H_4^\dagger D H_4 ,
\]
where \(D\) is the diagonal phase mask. The two bases are mutually unbiased when every element of \(U(D)\) has modulus \(1/\sqrt{4}=1/2\).

 \begin{table}[t]
\centering
\caption{The eight states prepared by Alice. The four $\mathcal{Z}$ states differ from their $\mathcal{X}$ partners by an extra $\pi$ phase on cores~$1$, $2$ and $3$ (equivalently, by an odd-parity element of $\mathbb{Z}_2^4$). Bob's $\mathcal{X}$-basis projection pattern $\vec{\phi}^B_{\mathcal{X}}=\{\pi,0,0,0\}$, which is realized here with a single PFPM, is equivalent up to a global phase to the symmetric pattern $\{0,\pi,\pi,\pi\}$ used at Alice for the $\ket{0_x}$ state.}
\label{tab:states}
\renewcommand{\arraystretch}{1.25}
\begin{tabular}{c l c}
\toprule
Basis & State & Phase pattern $\vec{\phi}^A$ \\
\midrule
$\mathcal{Z}$ & $\ket{0_z}=\tfrac{1}{2}\bigl(\ket{0}+\ket{1}+\ket{2}+\ket{3}\bigr)$ & $\{0,0,0,0\}$ \\
              & $\ket{1_z}=\tfrac{1}{2}\bigl(\ket{0}-\ket{1}+\ket{2}-\ket{3}\bigr)$ & $\{0,\pi,0,\pi\}$ \\
              & $\ket{2_z}=\tfrac{1}{2}\bigl(\ket{0}+\ket{1}-\ket{2}-\ket{3}\bigr)$ & $\{0,0,\pi,\pi\}$ \\
              & $\ket{3_z}=\tfrac{1}{2}\bigl(\ket{0}-\ket{1}-\ket{2}+\ket{3}\bigr)$ & $\{0,\pi,\pi,0\}$ \\
\midrule
$\mathcal{X}$ & $\ket{0_x}=\tfrac{1}{2}\bigl(\ket{0}-\ket{1}-\ket{2}-\ket{3}\bigr)$ & $\{0,\pi,\pi,\pi\}$ \\
              & $\ket{1_x}=\tfrac{1}{2}\bigl(\ket{0}+\ket{1}-\ket{2}+\ket{3}\bigr)$ & $\{0,0,\pi,0\}$ \\
              & $\ket{2_x}=\tfrac{1}{2}\bigl(\ket{0}-\ket{1}+\ket{2}+\ket{3}\bigr)$ & $\{0,\pi,0,0\}$ \\
              & $\ket{3_x}=\tfrac{1}{2}\bigl(\ket{0}+\ket{1}+\ket{2}-\ket{3}\bigr)$ & $\{0,0,0,\pi\}$ \\
\bottomrule
\end{tabular}
\end{table}

For the four-dimensional receiver, it is sufficient to apply a \(\pi\) phase shift to a single path. Choosing, without loss of generality,
\[
D_\pi=\mathrm{diag}(-1,1,1,1),
\]
one obtains
\[
U(D_\pi)
=H_4^\dagger D_\pi H_4
=
\frac{1}{2}
\begin{pmatrix}
1&-1&-1&-1\\
-1&1&-1&-1\\
-1&-1&1&-1\\
-1&-1&-1&1
\end{pmatrix}.
\]
Therefore
\[
\left|[U(D_\pi)]_{jk}\right|=\frac{1}{2}
\qquad
\text{for all } j,k\in\{1,2,3,4\}.
\]
This shows that the basis obtained with a single \(\pi\)-phase shift is mutually unbiased with respect to the original Hadamard analyzer basis. Hence, in \(d=4\), one independently driven phase modulator is sufficient to switch between the two measurement bases used in the protocol.

The argument above shows that, in $d=4$, a single phase modulator is sufficient to switch between the two MUBs at the level of the phase pattern. The corresponding hardware, one modulator on arm~$0$ and bare single-mode patches on the remaining three arms, introduces a small amplitude imbalance between the modulated and bare paths that contributes a mode-dependent term to the QBER. Consider the $\mathcal{Z}$-basis state prepared by Alice,
\[
\ket{0_z}=\tfrac{1}{2}\bigl(\ket{0}+\ket{1}+\ket{2}+\ket{3}\bigr),
\]
and assume arm~$0$ has power transmission $\eta_0=10^{-L/10}$, where $L$ is the modulator insertion loss in dB, while arms~$1$, $2$ and $3$ have $\eta_{1,2,3}\simeq 1$. Because the relevant object in the quantum state is the field amplitude rather than the optical power, the action of the lossy arm on the corresponding ket is multiplication by the amplitude transmission $\sqrt{\eta_0}$. Up to global phases absorbed into Bob's per-block calibration, the state at the input of the second 4C-MBS, conditional on no loss event, reads
\[
\ket{\psi_{\mathrm{MDL}}}
= \frac{1}{\sqrt{\eta_0+3}}\bigl(\sqrt{\eta_0},\,1,\,1,\,1\bigr),
\]
and the squared overlap with the ideal projector gives the mode-dependent-loss contribution to the QBER,
\begin{equation}
\mathrm{QBER}_{\mathrm{MDL}}(\eta_0)
\;=\;1\;-\;
\frac{\bigl(\sqrt{\eta_0}+3\bigr)^{\!2}}{4\,(\eta_0+3)}.
\label{eq:qber-mdl}
\end{equation}
By symmetry of the Sylvester--Hadamard transformation under sign flips of the equal-amplitude states, Eq.~\eqref{eq:qber-mdl} applies to every ideal MUB state of the protocol when arm~$0$ is the only lossy arm. Evaluated at the PFPM operating point ($L=0.45\,$dB, $\eta_0\simeq 0.901$), Eq.~\eqref{eq:qber-mdl} gives $\mathrm{QBER}_{\mathrm{MDL}}\simeq 5\times 10^{-4}$, two orders of magnitude below the measured baseline interferometric QBER of the receiver and therefore absorbed into Bob's per-block phase calibration together with the static phases of the three patch cords.

\section*{Acknowledgements}

This research was funded by ANID Anillo Project ATE250003, Fondo Nacional de Desarrollo Cient\'{\i}fico y
Tecnol\'ogico (FONDECYT) Grant No.\ 1260111, 1240746, 1240843, 1231940, ANID Millennium Science Initiative Program ICN17\textendash012, ANID AC3E CIA 250006, and Vinnova (project no. 2023-01358), the Wallenberg Center for Quantum Technologies (WACQT).

\bibliography{main.bib}

\end{document}